\documentclass[a4paper,11pt]{article}
\usepackage{jinstpub} 
\usepackage{subcaption}

\title{In-situ suppression of surface radon emanation using an HDPE barrier at the Stawell Underground Physics Laboratory (SUPL)}

\author[a,f,1]{R.R. Marcelo Gregorio,\note{Corresponding author.}}
\author[b,f]{K. Mintern-Lane,}
\author[b,f]{S. Rushbrook,}
\author[a,c,f]{A.G. McLean,}
\author[a,f]{L.J. Bignell,}
\author[d,f]{J.M.C. Brown,}
\author[a,f]{ G.J. Lane,}
\author[e]{ and N.J.C. Spooner}

\affiliation[a]{ Department of Nuclear Physics and Accelerator Applications, Australian National University, \\Garran Road, ACT 2601, Canberra, Australia}
\affiliation[b]{Stawell Underground Physics Laboratory, via Stawell Gold Mines, \\Reefs Road, Stawell, VIC 3380,  Australia}
\affiliation[c]{School of Physics, Chemistry and Earth Sciences, Adelaide University, \\North Terrace Campus, Adelaide, SA 5005, Australia }
\affiliation[d]{Department of Physics and Astronomy, Swinburne University of Technology\\ Hawthorn, VIC 3122, Australia}
\affiliation[e]{School of Mathematical and Physical Sciences, University of Sheffield, \\Hounsfield Road, Sheffield, South Yorkshire, S3 7RH, United Kingdom}
\affiliation[f]{ARC Centre of Excellence for Dark Matter Particle Physics, Australia}

\emailAdd{robert.gregorio@anu.edu.au}

\abstract{
Radon-induced backgrounds remain a limiting factor in low-background experiments conducted in underground laboratories. The enclosed nature of these laboratories can result in elevated radon levels, with emanation from the surrounding rock representing the primary source of radon. In this work, we present an in-situ study of radon emanation from the floor of the main experimental hall at the Stawell Underground Physics Laboratory (SUPL) and investigate the use of a high-density polyethylene (HDPE) barrier to suppress emanation. Surface emanation was measured using an AIRTHINGS Corentium Pro  passive detector before and after installation of the barrier. The equilibrium radon concentration within the surface emanation setup was reduced from $5840 \pm 50$~Bq\,m$^{-3}$ to $259 \pm 3$~Bq\,m$^{-3}$, corresponding to a suppression of $\sim$96\%. These results demonstrate the potential to retrofit radon barriers to suppress surface emanation under realistic underground conditions, supporting their use in planned low-background infrastructure at SUPL and future applications at other underground laboratories. }


\begin{document}
\maketitle
\flushbottom

\section{Introduction}
\label{sec:intro}

Many low-background experiments operate in deep underground laboratories, where the surrounding rock provides substantial shielding from cosmic radiation. However, the enclosed environment and large surface areas exposed to the surrounding rock can result in the accumulation of naturally occurring radon.
For example, radon concentrations exceeding $2000$~Bq\,m$^{-3}$ have been measured in mine tunnels surrounding the Kamioka Observatory underground laboratory \cite{pronost2020radon}. Radon and its progeny represent an important background for many low-background experiments \cite{aalbers2025dark, bo2025dark, aprile2025wimp, dolinski2019neutrinoless, poudel2026accelerating}. Of particular concern is the build-up of long-lived $^{210}$Pb following the plate-out of radon progeny onto material and detector surfaces, which can result in persistent backgrounds long after the initial radon exposure. Maintaining low-radon conditions within the underground laboratory throughout the storage, construction and operation of these experiments is crucial.

A range of approaches are employed to control radon in underground laboratories. These include active techniques such as dilution through ventilation with surface air and dedicated radon removal systems \cite{gregorio2024molecular, perez2022radon, cui2024environmental, street2018radon}. Passive approaches include radon-resistant barriers designed to prevent radon entering the laboratory environment, which are typically incorporated during the construction of the underground infrastructure. A comparatively unexplored approach is the application of radon barriers to existing infrastructure as a retrofit measure. This is particularly attractive because it further reduces the radon source term  and can complement existing active mitigation systems. Radon-resistant barriers are well established within the construction industry, with high-density polyethylene (HDPE) commonly used as a barrier material \cite{papp2015methods}. HDPE is commercially available, inexpensive and adaptable to existing infrastructure without substantial structural modification.

The Stawell Underground Physics Laboratory (SUPL) is the only active deep underground laboratory in the Southern Hemisphere and is located 1025~m underground within the Stawell Gold Mines in Victoria, Australia. The laboratory has an ambient radon concentration of  $\sim500$~Bq\,m$^{-3}$ under normal ventilation conditions. As a relatively new facility with further low-background infrastructure under development, SUPL provides an ideal environment to evaluate radon barriers.

In this paper, the experimental setup at SUPL to measure surface emanation is described in \autoref{sec:setup}, followed by details of the method used to evaluate radon suppression using an HDPE barrier in \autoref{sec:method}. The data analysis used to determine the surface emanation is presented in \autoref{sec:DataAnalysis}. Finally, the results and their implications for retrofitting low-background infrastructure are discussed in \autoref{sec:results}.


\section{In-situ surface emanation test setup}
\label{sec:setup}

\begin{figure}[htbp]
    \centering
    \includegraphics[width=15cm]{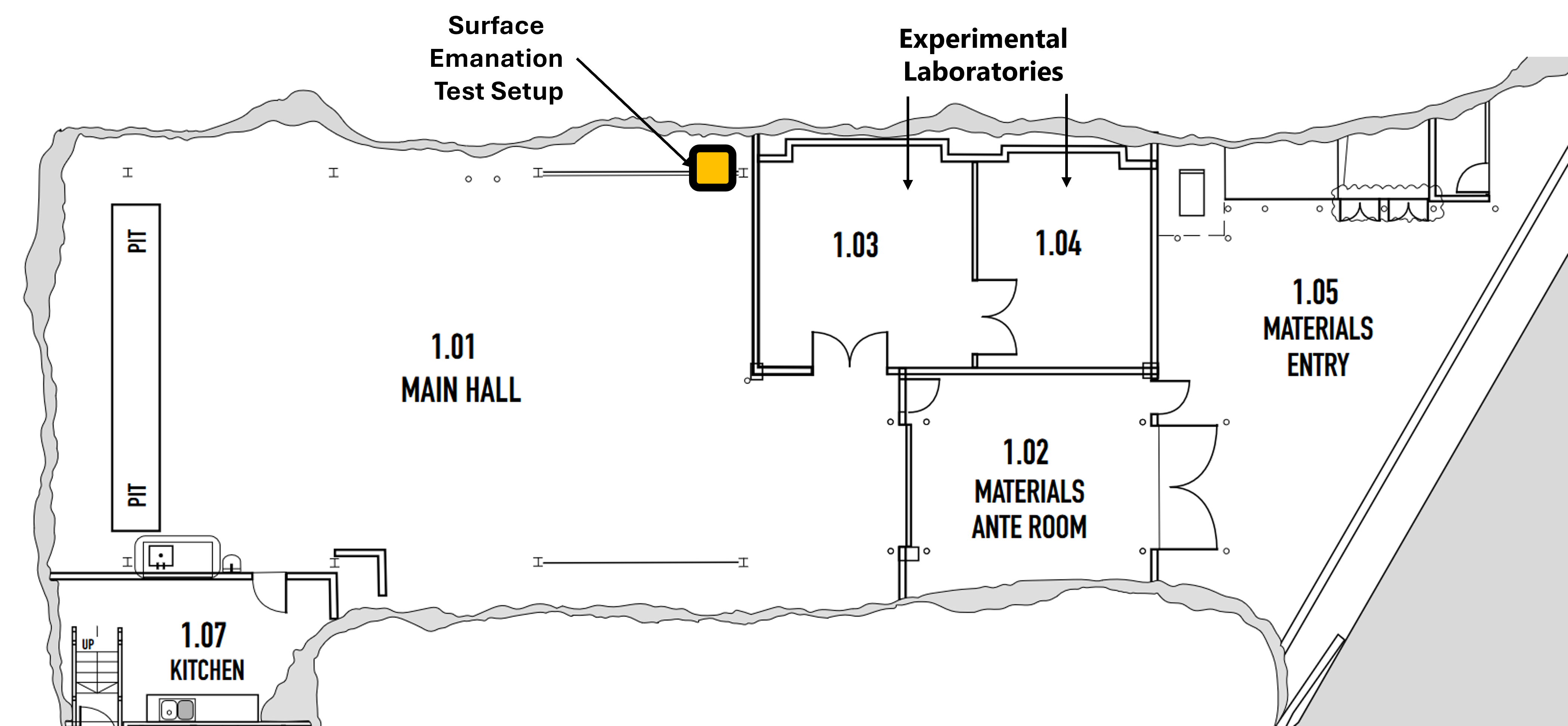}
\caption{Floor layout of SUPL showing the main experimental hall, experimental laboratories and the location of the surface emanation measurement.}
    \label{fig:SUPL_MAP}
\end{figure}

\autoref{fig:SUPL_MAP} shows the layout of SUPL. The cavern has a height of 12~m and a floor area of 10~m $\times$ 25~m, with an epoxy-coated floor and surrounding rock walls lined with Tekflex. The cavern houses the main experimental hall, two smaller experimental laboratories with a height of $\sim$5~m (1.03 and 1.04), and a materials ante room (1.02). Engineering studies have shown the composition of the surrounding rock is relatively uniform across the main experimental hall. Therefore, substantial spatial variation in surface emanation is not expected. The measurement location, indicated by the yellow square, was selected away from routine laboratory activities to minimise disturbance, while remaining near the region designated for the flagship detector, which is under construction at the time of writing \cite{zurowski2023status}.

The focus of this study is specifically on surface radon emanation from the epoxy floor of the main experimental hall. Floor emanation represents a more direct source of radon to infrastructure such as radon-reduced air tents, detector enclosures and cleanrooms. In contrast, emanation from the wider cavern surfaces is better addressed through laboratory ventilation.

\begin{figure}[htbp]
    \centering
    \includegraphics[width=13cm]{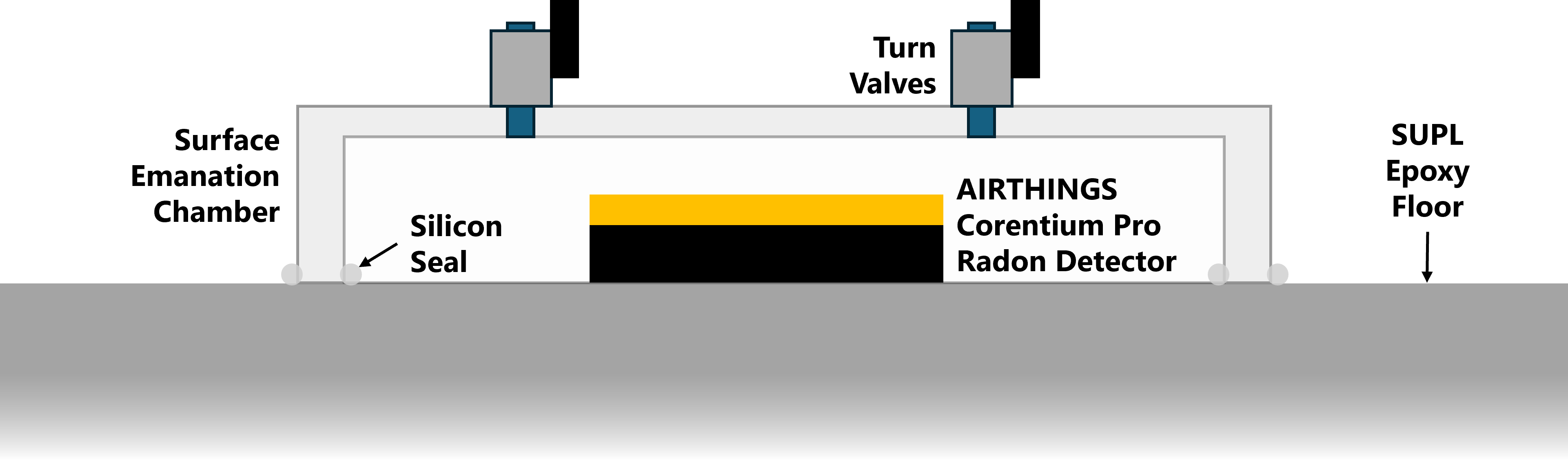}
    \caption{Schematic of the surface emanation test setup.}
    \label{fig:setup}
\end{figure}

\autoref{fig:setup} shows the surface emanation test setup installed on the epoxy floor of the main experimental hall at SUPL. The setup consists of an AIRTHINGS Corentium Pro passive radon detector housed within a $380 \times 260 \times 45$~mm$^3$ (internal dimensions) surface emanation chamber. To isolate radon emanating from the enclosed floor surface, the interface between the chamber walls and the SUPL epoxy floor was sealed with silicone. To minimise the contribution from ambient radon in the laboratory, two valves located on top of the chamber allow the volume to be purged with radon-reduced air before each emanation measurement.

\section{Measurement method}
\label{sec:method}

\begin{figure}[htbp]
    \centering
    \begin{subfigure}[t]{0.48\textwidth}
        \centering
        \includegraphics[height=5cm]{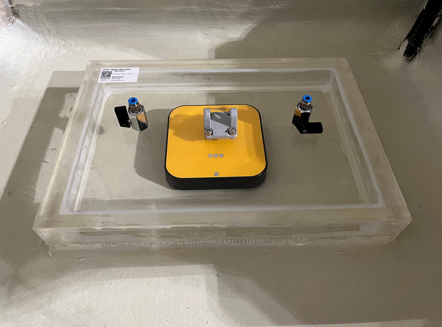}
        \caption{SUPL epoxy floor run.}
        \label{fig:SUPL_without}
    \end{subfigure}
    \hfill
    \begin{subfigure}[t]{0.48\textwidth}
        \centering
        \includegraphics[height=5cm]{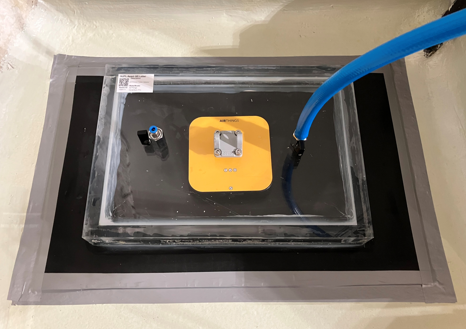}
        \caption{HDPE radon barrier run during purge.}
        \label{fig:SUPL_barrier}
    \end{subfigure}
    \caption{Photographs of the surface radon emanation measurements.}
    \label{fig:SUPL_floor}
\end{figure}

\autoref{fig:SUPL_floor} shows the two surface emanation measurements performed in the main experimental hall at SUPL: directly from the epoxy floor (\autoref{fig:SUPL_without}) and with the HDPE barrier installed (\autoref{fig:SUPL_barrier}). The two measurement configurations were conducted in separate runs using the same experimental setup at the same location. The barrier selected for this study was a 4~mm thick, commercially available general-purpose HDPE sheet supplied as a sample by \textit{Henan Okay Plastic Industry Co., Ltd}. The material is commercially priced at approximately USD~3.50 per kg. A generic, low-cost and readily available product was deliberately selected to assess whether substantial radon suppression could be achieved without the use of specialised low-background materials.

Before each measurement, the surface emanation chamber was sealed to the test surface using silicone sealant and allowed to set. For the SUPL epoxy floor measurement, the chamber was sealed directly to the floor. For the HDPE measurement, the chamber was sealed to the HDPE barrier, which itself was secured and sealed to the epoxy floor using high-strength adhesive tape. 
The chamber was purged for 24~hours using compressed air supplied from the surface, with a radon concentration of $\sim20$~Bq\,m$^{-3}$, to reduce the initial contribution from the higher ambient radon concentration within the underground laboratory. The two valves on the chamber were used as the purge inlet and outlet. Observation of flow from the outlet valve during purging also provided a qualitative check that the chamber remained sealed to the test surface. Following purging, both valves were closed and the chamber was left sealed to allow radon to accumulate. Each configuration was measured for 21~days, providing sufficient time to observe the approach towards secular equilibrium.

\section{Data analysis}
\label{sec:DataAnalysis}

To determine the surface emanation rate, it is first necessary to consider the behaviour of radon within the measurement chamber. Radon emanating from the enclosed surface accumulates within the chamber while simultaneously undergoing radioactive decay. Assuming a sealed chamber and a constant emanation rate, the radon concentration is described by

\begin{equation}
    C(t)
    =
    C_{\mathrm{eq}}
    +
    \left(C_0-C_{\mathrm{eq}}\right)
    \exp\left(-t/\tau_{\mathrm{Rn}}\right),
    \label{eq:radon_accumulation}
\end{equation}
where $C(t)$ is the radon concentration at time $t$, $C_0=20$~Bq\,m$^{-3}$ is the initial radon concentration assumed from the purge gas, $C_{\mathrm{eq}}$ is the radon concentration at secular equilibrium and $\tau_{\mathrm{Rn}}$ is the mean lifetime of $^{222}$Rn. At secular equilibrium, the activity of radon within the chamber is equal to the emanation rate from the enclosed surface. The absolute radon emanation, $E$, can be determined from the fitted equilibrium concentration, $C_{\mathrm{eq}}$, according to

\begin{equation}
    E = C_{\mathrm{eq}}V,
    \label{eq:absolute_emanation}
\end{equation}

where $V$ is the free internal volume of the measurement chamber. The chamber has an internal volume of 4.45~L, of which 0.59~L is displaced by the AIRTHINGS Corentium Pro, giving a free internal volume of $V=3.86$~L. The resulting absolute emanation rate, $E$, is expressed in Bq, where 1~Bq corresponds to the release of one $^{222}$Rn atom from the surface per second at secular equilibrium. Surface emanation rates are typically normalised to the exposed surface area, $A$. The chamber footprint corresponds to an exposed floor area of $A=0.099$~m$^2$. The surface-normalised radon emanation rate, $E_A$, expressed in Bq\,m$^{-2}$, is given by

\begin{equation}
    E_A
    =
    \frac{E}{A}
    =
    \frac{C_{\mathrm{eq}}V}{A}.
    \label{eq:surface_emanation}
\end{equation}

The Corentium Pro reports radon concentrations in hourly measurement cycles. As a passive diffusion-based detector, its short-timescale response is limited by the diffusion of radon into the active detector volume \cite{airthings2021corentium}. The hourly measurements were resampled into 24-hour intervals to reduce short-timescale variability. \autoref{eq:radon_accumulation} was fitted to the radon concentrations over the measurement period for both the SUPL epoxy floor and HDPE configurations. The fitted $C_{\mathrm{eq}}$ was subsequently used to calculate the absolute emanation rate, $E$, and surface-normalised emanation rate, $E_A$, using \autoref{eq:absolute_emanation} and \autoref{eq:surface_emanation}, respectively.

\section{Results and discussion}
\label{sec:results}

\begin{figure}[htbp]
    \centering
    \begin{subfigure}[t]{0.48\textwidth}
        \centering
        \includegraphics[height=5.5cm]{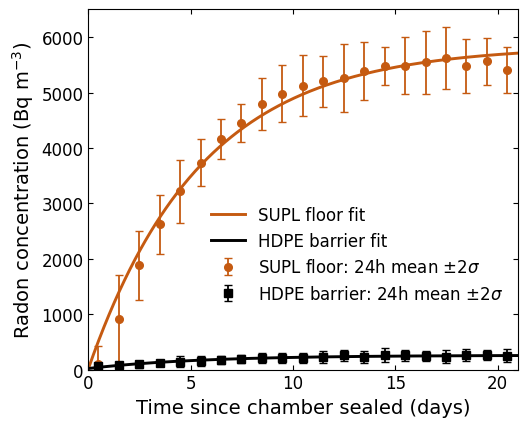}
        \caption{SUPL epoxy floor and HDPE barrier comparison.}
        \label{fig:SUPL_comparison}
    \end{subfigure}
    \hfill
    \begin{subfigure}[t]{0.48\textwidth}
        \centering
        \includegraphics[height=5.5cm]{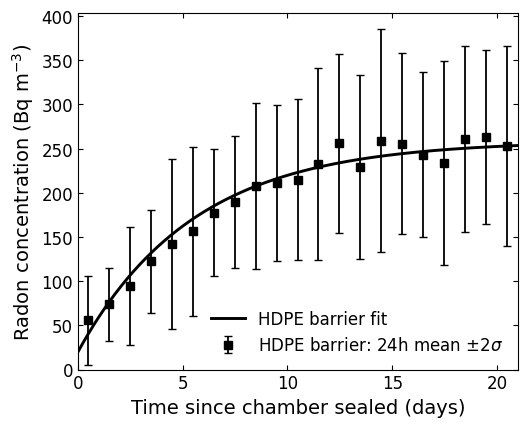}
        \caption{HDPE barrier on zoomed scale.}
        \label{fig:SUPL_zoom}
    \end{subfigure}
    \caption{Radon concentration as a function of time since the chamber was sealed.}
    \label{fig:SUPL_data}
\end{figure}

\autoref{fig:SUPL_comparison} shows the radon concentration as a function of time since the surface emanation chamber was sealed, with the SUPL epoxy floor run shown in orange and the HDPE barrier run in black. The data points represent 24-hour mean concentrations, with error bars corresponding to $\pm2\sigma$ of the hourly measurements within each interval. The relatively large spread reflects the limited short-timescale precision of the Corentium Pro. The solid lines represent fits to \autoref{eq:radon_accumulation}, from which the secular equilibrium concentration, $C_{\mathrm{eq}}$, was determined. \autoref{fig:SUPL_zoom} shows the HDPE barrier measurement separately on a zoomed radon concentration scale.

\autoref{tab:emanation_results} shows the calculated equilibrium concentrations, absolute emanation and surface-normalised emanation. A clear suppression of radon emanation is observed following installation of the HDPE barrier. The fitted secular equilibrium concentration decreased from $5840 \pm 50$~Bq\,m$^{-3}$ to $259 \pm 3$~Bq\,m$^{-3}$, corresponding to a suppression of approximately 96\%, or a factor of 23.

\begin{table}[htbp]
    \centering
    \caption{Radon emanation results for the SUPL epoxy floor and HDPE barrier configurations.}
    \label{tab:emanation_results}
    \begin{tabular}{lccc}
        \hline
        Measurement &
        $C_{\mathrm{eq}}$ (Bq\,m$^{-3}$) &
        $E$ (Bq) &
        $E_A$ (Bq\,m$^{-2}$) \\
        \hline
        SUPL epoxy floor & $5840 \pm 50$ & $22.5 \pm 0.2$ & $227 \pm 2$ \\
        HDPE barrier    & $259 \pm 3$   & $1.00 \pm 0.01$ & $10.1 \pm 0.1$ \\
        \hline
    \end{tabular}
\end{table}

These results demonstrate the potential of retrofitted radon barriers as a passive approach to suppressing surface emanation under realistic underground conditions. By reducing radon at its source, surface barriers can complement the active ventilation and radon-reduced air systems commonly employed in underground laboratories. This could enable lower radon concentrations to be achieved or reduce the high ventilation rates required to maintain a target radon concentration.

The use of a commercial HDPE barrier provides a comparatively low-cost solution that can be retrofitted without requiring substantial modification. This makes the approach applicable to existing infrastructure, such as radon-reduced air tents and detector enclosures, and directly relevant to the ambition to convert one of the experimental laboratories at SUPL into a dedicated low-radon environment. The need for this is further motivated by the growing range of Australian activities for which radon represents an important background that could benefit from SUPL’s infrastructure, including rare-event physics searches such as SABRE South \cite{fu2026measurement}, CYGNUS-Oz \cite{vahsen2020cygnus} and ORGAN (Oscillating Resonant Group AxioN) \cite{mcallister2017organ} , an international underground radiobiology study \cite{fedele2025broadening}, quantum technology research through the Cryogenic Experimental Laboratory for Low-background Australian Research (CELLAR) Facility \cite{campbell2026search}, and ultra-sensitive material assay through the Detecting Radon Emanation and Mitigating Radon (DREAMR) Facility \cite{marcelo2026direct}. In future, combining passive surface barriers with an active radon removal system could further suppress radon backgrounds and support the development of world-class low-background infrastructure at SUPL.

\section{Conclusion}
\label{sec:conclusion}

In this paper, an in-situ study of surface radon emanation from the floor of the main experimental hall at the Stawell Underground Physics Laboratory (SUPL) is presented. The effectiveness of a high-density polyethylene (HDPE) barrier for suppressing this emanation was investigated using a surface emanation chamber with an AIRTHINGS Corentium Pro radon detector. The surface emanation was measured from the SUPL epoxy floor with and without a commercially available 4~mm HDPE barrier installed. It was found that the surface radon emanation was reduced from $227 \pm 2$ to $10.1 \pm 0.1$~Bq m$^{-2}$, corresponding to a suppression of $\sim96\%$, or a factor of 23. These results demonstrate that retrofitted surface barriers can provide substantial passive suppression of radon emanation under realistic underground conditions and when combined with active radon mitigation, offer a practical and low-cost approach to reducing radon backgrounds in future low-background infrastructure at SUPL and other underground laboratories.

\acknowledgments

The authors would like to: (1) acknowledge that this work was supported through the Australian Research Council Centre of Excellence for Dark Matter Particle Physics Grant (CE200100008), and (2) thank the Australian Radiation Protection and Nuclear Safety Agency (ARPANSA) Radiation Health Services Branch for the custom surface emanation chamber and loan of the AIRTHINGS Corentium Pro radon detector utilised in this work.


\bibliographystyle{JHEP}
\bibliography{biblio.bib} 

\end{document}